\documentclass[10pt]{article}
 
\usepackage{amsmath}
\usepackage{amsthm}
\usepackage{amssymb}
\usepackage{amscd}
\usepackage{epsf}
\usepackage{rotating}
\usepackage{wrapfig}

\chardef\bslash=`\\ 

\makeatletter
\def\verbatim{\interlinepenalty\@M \@verbatim
  \leftskip\@totalleftmargin\advance\leftskip2pc
  \frenchspacing\@vobeyspaces \@xverbatim}
\makeatother
\def\1I{\relax{\rm 1\kern-.25em \rm l}} 

\newcommand{\be}{\begin{equation}}
\newcommand{\ee}{\end{equation}}

\def\href#1#2{#2}

\begin{document}

\thispagestyle{empty}
\rightline{hep-th/0312096}
\vspace{2truecm}

\begin{center}

{\bf \Large $E_{11}$ and Spheric Vacuum Solutions of Eleven- and Ten 
Dimensional Supergravity Theories} 
\end{center}

\vspace{.5truecm}

\newcounter{Institut}
\vspace{1.5truecm}
\centerline{\bf Igor Schnakenburg\refstepcounter{Institut}\label{Inst_Igor}{}$^{*_{\theInstitut}}$
            and
               Andr\'e Miemiec\refstepcounter{Institut}\label{Inst_Andre}{}$^{*_{\theInstitut}}$}

\vskip1cm
\parbox{.9\textwidth}{
\parbox{.4\textwidth}{
\centerline{$^{*_{\ref{Inst_Igor}}}$~  Racah Institute of Physics}
\centerline{                           The Hebrew University}
\centerline{                           Jerusalem 91904}
\centerline{                           Israel}
\centerline{                           igorsc@phys.huji.ac.il}
}\hfill
\parbox{.4\textwidth}{
\centerline{$^{*_{\ref{Inst_Andre}}}$~ Institut f\"ur Physik}
\centerline{                           Humboldt Universit\"at}
\centerline{                           D-12489 Berlin, Newtonstr. 15}
\centerline{                           Germany}
\centerline{                           miemiec@physik.hu-berlin.de}
}
}

\vspace{1.0truecm}
\begin{abstract}
 \noindent
In view of the newly conjectured Kac-Moody symmetries of supergravity 
theories placed in eleven and ten dimensions, the relation between 
these symmetry groups and possible compactifications are examined. 
In particular, we identify the relevant group cosets that parametrise
the vacuum solutions of $AdS\times S$ type.
\end{abstract}
\bigskip \bigskip

\newpage

\section{Introduction}
Recently a class of infinite dimensional groups has been
conjectured to play a role in the formulation of supergravity
(superstring) theories. The generating algebras are Kac-Moody 
algebras (KMA). The relation between these algebras and uncompactified 
(super) gravity theories has first been suggested in [1]. Afterwards it 
was shown that almost all (super)gravities admit a description in terms 
of non-linear realisation  of groups generated by subsets of the 
generators of a KMAs [2,3,4,5]. The low level roots of these algebras 
were shown to be in one-to-one correspondence with the field contents 
of these theories in [6]. Further properties of these particular 
Kac-Moody algebras were presented in [7]. Relations to string theory were 
examined in [8,9].
\par\noindent
The question we focus on here is to identify those subsets of generators 
that generate the isometries of solutions of the supergravity under 
consideration. We will show how some known symmetry algebras occur from 
$E_{11}$. In the next two sections we want to discuss the situation in 
eleven and ten dimensions.

\section{Possible vacua of $D=11$ supergravity}

The nonlinear realisations under consideration are cosets $G/H$ where $G$ 
is generated by a KMA and $H$ is generated by all those generators that 
are left invariant under a modified Cartan involution, called 'temporal' 
involution. It was  presented in [8] and defined in the following way. 
The algebra $\mathfrak{g}$ of $G$ can be decomposed into a sum 
$\mathfrak{g}= \Delta^-\oplus \,{\rm CSA}\,\oplus \Delta^+$, where 
$\Delta^\pm$ are called positive/negative roots and CSA denotes the 
Cartan subalgebra. On the elements of the Cartan subalgebra one defines 
an action by
\be\label{TI on CSA}
   H_a \rightarrow -H_a,
\ee
and on the simple positive and negative roots by
\be
   E_\alpha \leftrightarrow -\epsilon_{\alpha} E_{-\alpha},
\ee
where $\alpha$ is parametrising any set of simple positive roots, and
$\epsilon_\alpha$ is defined according to the number of indices of the 
relevant generator that is pointing in the time-like direction, i.e.
it is taken to be +1 if the number of indices pointing in the time 
direction ('1') is even, and -1 otherwise. The action on the simple 
roots extends to an action on all roots.
The linear combination $E_\alpha - \epsilon_\alpha E_{-\alpha}$ is left 
invariant under this involution. The coset G/H is obviously parametrised 
by the linear combinations $E_{\alpha} + \epsilon_\alpha E_{-\alpha}$ 
and the elements of the Cartan subalgebra.
\par\noindent
An important example are $A_{d-1}$ algebras with generators $K^a{}_b$. 
The linear combination $E_\alpha - \epsilon_\alpha E_{-\alpha}$ singles 
out the antisymmetric generators $J_{ab}= K_{ab}-K_{ba}$. Here the indices 
are pulled down with the flat metric $\eta_{ab}$, which we have chosen to 
be $\eta_{ab}={\rm diag}(-,+\ldots +)$. These are the generators of the 
Lorentz group $SO(1,d-1)$\footnote{The 'pure' Cartan involution leads to 
the invariant subgroup $SO(d)$.}.\\
\par\noindent
In the non-linear realisation $G/H$ we take a group element $g\in G$ to 
transform as
\be
   g\to g_0 \, g\, h^{-1},
\ee
with $h\in H$. If the Cartan form $g^{-1}dg$ is 
enhanced by the Lorentz connection term $\omega$ [1,10]
\be
   \mathcal{V} = g^{-1}dg -\omega,
\ee
then $\mathcal{V}$ transforms covariantly under $H$
\be
   \mathcal{V} \to h\mathcal{V} h^{-1}.
\ee
In particular, if $H$ is the Lorentz group (see above) then the 
corresponding Cartan form $\mathcal{V}$ is Lorentz covariant [10].\\
\par\noindent
In [1] it was conjectured that eleven dimensional supergravity admits a 
description as a nonlinear realisation of the Kac-Moody algebra 
$E_{11}$, whose Dynkin diagram is depicted in {\it Figure A} below.
\\[5ex]
\parbox{\textwidth}{
\begin{picture}(-150,0)(-50,30)
 \put(10,20){\circle{10}}\put(8,5){1}
 \put(15,20){\line(1,0){10}}
 \put(30,20){\circle{10}}\put(28,5){2}
 \put(35,20){\line(1,0){10}}
 \put(50,20){\circle{10}}\put(48,5){3}
 \put(55,20){\line(1,0){10}}
 \put(70,20){\circle{10}}\put(68,5){4}
 \put(75,20){\line(1,0){10}}
 \put(90,20){\circle{10}}\put(88,5){5}
 \put(95,20){\line(1,0){10}}
 \put(110,20){\circle{10}}\put(108,5){6}
 \put(115,20){\line(1,0){10}}
 \put(130,20){\circle{10}}\put(128,5){7}
 \put(135,20){\line(1,0){10}}
 \put(150,20){\circle{10}}\put(148,5){8}
 \put(155,20){\line(1,0){10}}
 \put(170,20){\circle{10}}\put(168,5){9}
 \put(175,20){\line(1,0){10}}
 \put(190,20){\circle{10}}\put(188,5){10}

 \put(150,25){\line(0,1){10}}
 \put(150,40){\circle{10}}\put(146,47){$R^3$}
\end{picture}
}\\
\vspace{.5cm}\\
\centerline{\it Figure A}
\noindent
This algebra obviously contains the subalgebra $A_{10}$, which 
in turn contains the subalgebra ${\mathfrak{so}}(1,10)$. It is the 
corresponding Lorentz group $SO(1,10)$, that becomes local ($H$) 
in the nonlinear realisation $G/H$. The particular Lorentz signature is 
a first consequence of the modified Cartan involution. 
\par\noindent
It is convenient to split representations of $E_{11}$ into 
representations of $A_{10}$ since in the latter it is simple to find
the Lorentz subalgebra. This analysis of splitting representations
with respect to their $A_n$ subalgebra has been carried out in [5,6] 
and we briefly give the results. According to the number the simple 
root $R^{9\,10\,11}$ appears in the non-simple roots of $E_{11}$ the 
representations of $A_{10}$ are organised in levels. At level zero of 
the decomposition the adjoint representation of $A_{10}\sim SL(11)$ is 
reproduced, while at level 1, 2, and 3 the generators 
\be\label{11d R3 R6 R8,1}
   R^{a_1a_2a_3}, \quad R^{a_1\ldots a_6}, \quad R^{a_1\ldots a_8,b}\quad\quad
   a_1,\ldots a_8,b \,=\,1\ldots 11
\ee
occur, respectively.
\par\noindent
We next want to split the $A_{10}$ subalgebra into subalgebras 
$A_6$ and $A_3$ by deleting the fourth node of the $A_{10}$ subalgebra
from the right. This corresponds to a split of $SL(11)$ into $SL(4)$ 
and $SL(7)$ where both subspaces completely decouple
\be\label{mixing term}
   [ K^{\hat a}{}_{\hat b},\, K^i{}_j]\equiv 0,
\ee 
(the unhatted $K^i{}_j$ lives in seven dimensions and the hatted 
one in four). The reason for these special values will become obvious 
in due course. We note that it is consistent with the commutation 
relations of $E_{11}$ to couple the level 1 generator to the left side 
of the subalgebra $A_3$, see {\it Figure B}. $A_4\oplus A_6$ 
is a proper, finite dimensional, and decomposable subalgebra of $E_{11}$.
\\[5ex]
\parbox{\textwidth}{
\begin{picture}(-150,0)(-50,30)
 \put(10,20){\circle{10}}\put(8,5){1}
 \put(15,20){\line(1,0){10}}
 \put(30,20){\circle{10}}\put(28,5){2}
 \put(35,20){\line(1,0){10}}
 \put(50,20){\circle{10}}\put(48,5){3}
 \put(55,20){\line(1,0){10}}
 \put(70,20){\circle{10}}\put(68,5){4}
 \put(75,20){\line(1,0){10}}
 \put(90,20){\circle{10}}\put(88,5){5}
 \put(95,20){\line(1,0){10}}
 \put(110,20){\circle{10}}\put(108,5){6}
 \put(150,20){\circle{10}}\put(148,5){8}
 \put(155,20){\line(1,0){10}}
 \put(170,20){\circle{10}}\put(168,5){9}
 \put(175,20){\line(1,0){10}}
 \put(190,20){\circle{10}}\put(188,5){10}

 \put(150,25){\line(0,1){10}}
 \put(150,40){\circle{10}}\put(146,47){$R^3$}
\end{picture}
}\\
\vspace{.5cm}\\
\centerline{\it Figure B}
\noindent
This includes that the 3-form generator can merely take values in the 
four dimensional subspace $A_3$ it couples to. 
We check explicitly that the 3-form potential enhances the 
aforementioned $SL(4)$ to $SL(5)$ by taking the commutation relations of 
$R^{\hat a_1\hat a_2\hat a_3}$ with $K^{\hat a}{}_{\hat b}$ 
written in terms of new generator introduced by:
\be\label{SL5_1}
   K^0{}_{\hat c} = \frac{1}{3!}\epsilon_{\hat a_1\hat a_2\hat a_3\hat c} 
   R^{\hat a_1\hat a_2\hat a_3},\quad
   K^{\hat c}{}_0 = \frac{1}{3!}\epsilon^{\hat a_1\hat a_2\hat a_3\hat c} 
   R_{\hat a_1\hat a_2\hat a_3} .
\ee
We find
\be\label{SL5_2}
   [K^{\hat a}{}_{\hat b},\, K^0{}_{\hat c}] = -\delta^{\hat a}_{\hat c}\, 
     K^0{}_{\hat b} \,+\,\delta^{\hat a}_{\hat b}\, K^0{}_{\hat c},\quad
   [K^{\hat a}{}_{\hat b},\, K^{\hat c}{}_0] = \delta^{\hat c}_{\hat b}\, 
   K^{\hat a}{}_0 \,-\,\delta^{\hat a}_{\hat b}\, K^{\hat c}{}_0.
\ee
Taking out a trace part by $\hat K^{\hat a}{}_{\hat b}\,=\,
K^{\hat a}{}_{\hat b}-\frac{1}{3}\,\delta^{\hat a}_{\hat b}\,
\sum_{\hat c=8}^{11} K^{\hat c}{}_{\hat c}$ we obtain 
\be\label{SL5_3}
   [\hat K^{\hat a}{}_{\hat b},\, K^0{}_{\hat c}] = -
    \delta^{\hat a}_{\hat c}\, K^0{}_{\hat b},\quad 
   [\hat K^{\hat a}{}_{\hat b},\, K^{\hat c}{}_0] = -
    \delta^{\hat c}_{\hat b}\, K^{\hat a}{}_0,
\ee
which provide the correct commutation relations to form the algebra 
of $SL(5)$ if we also add the new diagonal element
\be
   K^0{}_0 = \frac{1}{3}\sum_{\hat{a}=8}^{11} K^{\hat a}{}_{\hat a} .
\ee
The typical shape of the Dynkin diagram of $SL(5)$ can be obtained from 
the simple, positive roots (labels as on the right hand side of 
{\it Figure B})
\be
   E_{R^3} = K^0{}_8, \, E_8 = K^8{}_9, \, E_9=K^9{}_{10},\, 
    E_{10}=K^{10}{}_{11},
\ee
and the elements of the CSA given by
\be
   H_{R^3}= -2K^8{}_8- K^9{}_9-K^{10}{}_{10}-K^{11}{}_{11},\, 
   H_8 = K^8{}_8-K^9{}_9-2K^{10}{}_{10}- 2K^{11}{}_{11},
\ee
\be
   H_9 = K^9{}_9 - K^{10}{}_{10}- 2K^{11}{}_{11},\,
   H_{10}= K^{10}{}_{10}-K^{11}{}_{11}.
\ee

\par\noindent
In the full non-linear realisation of $E_{11}$ [2,3,10] one finds the 
covariant quantity belonging to the generator 
$R^{\hat a_1\hat a_2\hat a_3}$ by calculating the Cartan form and taking 
the closure with the conformal group. The resulting object is 
$F_{\hat a_1\hat a_2\hat a_3\hat a_4}=4 \partial_{[\hat a_1}A_{\hat a_2\hat
  a_3\hat a_4]}$ and can be 
identified with the field strength of eleven dimensional 
supergravity. Since the splitting of the algebra allows dependence on
4 dimensions only, we have to conclude that this field strength is
proportional to $\epsilon_{\hat a_1\hat a_2\hat a_3\hat a_4}$ (Freund-Rubin ansatz). 
\par\noindent
The method of nonlinear realisations applied to supergravity theories 
requires doubling of the fields, i.e. apart from the 3-form generator
which generates the 3-form gauge potential, one also introduces the 
magnetic dual in form of a 6-form generator belonging to a 6-form 
gauge potential. The Lorentz covariant field equation that reduces the 
doubled degrees of freedom is a generalised self-duality relation
\be
   F^{a_1a_2a_3a_4}= \frac{1}{7!}\epsilon^{a_1\ldots a_{11}}
   F_{a_5\ldots a_{11}}.
\ee 
This equation indicates that the field strength of the 6-form gauge 
potential is living in the space orthogonal to the chosen 4 dimensions. 
To this 6-form potential belongs the generator $R^{c_1\ldots c_6}$. 
We are forced to conclude, that the 6-form generator which is redundant 
in the Dynkin diagram of $E_{11}$ since it does not correspond to a 
simple root must be included by hand due to the non-trivial equation of
motion and the requirement that it does not vanish. This poses an
algebraic puzzle since there is no root of $E_{11}$ which might enlarge 
the $A_6$ subalgebra, while keeping the $SL(5)$ algebra unaffected
\footnote{We thank A.Kleinschmidt for discussion on that subject.}.  
However, introducing 
\begin{eqnarray}\label{6FormMomentum}
   K^{-1}{}_i &=& \frac{1}{6!}\epsilon_{ii_1\ldots i_6} R^{i_1\ldots i_6},\\
   K^{i}{}_{-1} &=& \frac{1}{6!}\epsilon^{ii_1\ldots i_6} R_{i_1\ldots i_6},
   \quad i\,=\,1,\ldots,7
\end{eqnarray}
and using the commutation relations of $E_{11}$ on this seven dimensional
subspace (indices $1,\ldots,7$), we find analogous formulas to 
(\ref{SL5_1} - \ref{SL5_3}). This suggests coupling of the 6-form 
generator to the seven-dimensional subspace. In fact, there is a root 
$\lambda$ of $E_{11}$ which is orthogonal at least to the $A_3$ 
subalgebra (labels 8, 9, 10) and leads to an extension of $A_6$ to 
$A_7$. It reads: 
\be
   \lambda = \alpha_2\,+\,2\alpha_3\,+\,3\alpha_4\,+\,4\alpha_5\,+\,
            5\alpha_6\,+\,6\alpha_7\,+\,6\alpha_8\,+\,4\alpha_9\,+\,
            2\alpha_{10}\,+\,2\alpha_{11}.
\ee
and is depicted below.
\pagebreak
\\[5ex]
\parbox{\textwidth}{
\begin{picture}(-150,0)(-50,30)
\put(10,25){\line(0,1){10}}
 \put(10,40){\circle{10}}\put(6,47){$R^6$}

 \put(10,20){\circle{10}}\put(8,5){1}
 \put(15,20){\line(1,0){10}}
 \put(30,20){\circle{10}}\put(28,5){2}
 \put(35,20){\line(1,0){10}}
 \put(50,20){\circle{10}}\put(48,5){3}
 \put(55,20){\line(1,0){10}}
 \put(70,20){\circle{10}}\put(68,5){4}
 \put(75,20){\line(1,0){10}}
 \put(90,20){\circle{10}}\put(88,5){5}
 \put(95,20){\line(1,0){10}}
 \put(110,20){\circle{10}}\put(108,5){6}
 \put(150,20){\circle{10}}\put(148,5){8}
 \put(155,20){\line(1,0){10}}
 \put(170,20){\circle{10}}\put(168,5){9}
 \put(175,20){\line(1,0){10}}
 \put(190,20){\circle{10}}\put(188,5){10}
\end{picture}
}\\
\vspace{.5cm}\\
\centerline{\it Figure C}
\noindent
\par\noindent
The simultaneous presence of gauge generators $R^3$ and $R^6$ 
(as required by the equation of motion) both of which enhance the 
relevant $A_3$ and $A_6$ subalgebras of $A_{10}$ can only be consistent 
with the algebraic structure of $E_{11}$ if commutation relations 
between elements of either algebra effectively vanish. This 
blindfolding of $E_{11}$ indeed occurs due to the field which is 
associated with the generator $R^{8,1}$. As shown, the vanishing of 
``mixing terms'' is trivial for the generators $\hat K^{\hat a}{}_{\hat b}$ and 
$K^i{}_j$, see \eqref{mixing term}. The only non-trivial ``mixing'' 
commutator  is
\be
   [R^{\hat a_1\hat a_2\hat a_3},\, R^{i_1\ldots i_6}] \sim_{E_{11}} 
   R^{i_1\ldots i_6[\hat a_1\hat a_2,\hat a_3]},
\ee
which is non-vanishing in $E_{11}$ (and therefore $A_4 \oplus A_7$ is not 
a subalgebra of $E_{11}$). The generator on the right hand side is of 
course the level $3$ generator from equation \eqref{11d R3 R6 R8,1}, 
antisymmetric in eight indices. We thus realise that the corresponding 
field cannot exist without destroying the split of $SL(11)$ caused by 
deleting the 7th node as we have done above. So we are provided with 
a natural cut-off for the roots of $E_{11}$\footnote{This generator 
tends to be a latent source for troubles since its interpretations as 
a gravity dual is not fully understood.}. One can use the Jacobi 
identity to show that all higher roots of $E_{11}$ cannot be present 
after this split since
\be
   [R^6,\, R^6]\, =\, [R^6,\,[R^3_1,\, R^3_2]] = [[R^6,\, R^3_1],\, R^3_2] 
   + [R^3_1,\, [R^6,\,R^3_2]] \equiv 0.
\ee
This argument shows that one can argue for an ``effective subalgebra'' 
$SL(8) \oplus SL(5)$ in $E_{11}$. This process is depicted in 
{\it Figure D}.\\[5ex]
\parbox{\textwidth}{
\begin{picture}(-150,0)(-50,30)
\put(10,25){\line(0,1){10}}
 \put(10,40){\circle{10}}\put(6,47){$R^6$}

 \put(10,20){\circle{10}}\put(8,5){1}
 \put(15,20){\line(1,0){10}}
 \put(30,20){\circle{10}}\put(28,5){2}
 \put(35,20){\line(1,0){10}}
 \put(50,20){\circle{10}}\put(48,5){3}
 \put(55,20){\line(1,0){10}}
 \put(70,20){\circle{10}}\put(68,5){4}
 \put(75,20){\line(1,0){10}}
 \put(90,20){\circle{10}}\put(88,5){5}
 \put(95,20){\line(1,0){10}}
 \put(110,20){\circle{10}}\put(108,5){6}
 \put(150,20){\circle{10}}\put(148,5){8}
 \put(155,20){\line(1,0){10}}
 \put(170,20){\circle{10}}\put(168,5){9}
 \put(175,20){\line(1,0){10}}
 \put(190,20){\circle{10}}\put(188,5){10}

 \put(150,25){\line(0,1){10}}
 \put(150,40){\circle{10}}\put(146,47){$R^3$}
\end{picture}
}\\
\vspace{.5cm}\\
\centerline{\it Figure D}
\noindent
Of course, this $A_7\oplus A_4$ algebra is finite.\\

\par\noindent
In the following we concentrate on the local subalgebras that arise via
the temporal involution as was indicated in the opening of this paragraph.
Since the $E_{11}$ algebra is assumed to be given in its maximal 
non-compact form, so is its $A_4\sim SL(5)$ subalgebra. We thus identify
$SO(5)$ as the compact subalgebra which is taken to be local in the 
relevant coset $SL(5)/SO(5)$. This coset is parametrised by the fields 
of the theory which only live in four dimensions spanned by the $A_3$ 
subalgebra of $A_{10}$, i.e. by the (traceless) generators 
$K_{\hat a\hat b}+ K_{\hat b\hat a}$ and the linear combination 
$R^{\hat a_1\hat a_2\hat a_3} + R_{\hat a_1\hat a_2\hat a_3}$ as well as 
the elements of the Cartan subalgebra given by appropriate linear 
combinations of $K^{\hat a}{}_{\hat a}$ ($\hat a=8,\ldots,11$). 
The local subgroup is spanned by the generators 
\be
   K_{\hat a\hat b}-K_{\hat b\hat a}\quad {\rm and}\quad 
      R^{\hat a_1\hat a_2\hat a_3}-R_{\hat a_1\hat a_2\hat a_3},
\ee
with minus signs since none of the indices is time-like. 
It has been noted in [11,8,3] 
that a generator coupling to the far left side of an $A_{d-1}$ 
subalgebra has the same Lorentz transformation properties as the 
momentum generator. This is precisely the situation we encountered 
in eq.~(\ref{SL5_1}).
In particular we have to apply the temporal involution to the 
generator $\hat P_{\hat a}$. Still, since none of the indices is 
time-like we find the local subalgebra to be spanned by generators
\footnote{$P_a$ denotes a generator, and is not the same as 
$\eta_{ab} P^b$.}
\be\label{SO(5) = SO(4)+ P}
   J_{\hat a\hat b} = K_{\hat a\hat b}-K_{\hat b\hat a},\quad {\rm and} 
   \quad P_{\hat c} = \hat P_{\hat c} - \eta_{\hat c\hat d}
   \hat P^{\hat d},
\ee
where indices are shovelled about with $\eta_{\hat a\hat b}= (+,+,+,+)$. 
The generators $J_{\hat a\hat b}$ correspond to the usual Lorentz 
generators in the 4 dimensional subalgebra of $A_3\sim SL(4)$. In view 
of the last equation \eqref{SO(5) = SO(4)+ P} one can further identify
the momentum generator with a generator of the Lorentz algebra in a new
direction $\#$
\be
   P_{\hat a} \equiv J_{\hat a\,\#}.
\ee
This momentum generator can now be taken to parametrise the coset
$\frac{SO(5)}{SO(4)}$, describing the previously discussed 
reduction of the aforementioned local $SO(5)$ symmetry to 
$SO(4)$ (pure Lorentz symmetry). This coset is well-known to define 
the 4-sphere $S^4$ and indicates that $E_{11}$ naturally contains
the part of the vacuum solution that is given by the 4-sphere.
\par
This is not the whole discussion since we have to deal with the left 
hand side of the diagram in {\it Figure D}. Again the node generating
the 6-form potential behaves exactly like a momentum operator of this 
subspace. In particular, the generator corresponding to the simple 
positive root on the far left of $A_7$ can in seven dimensions be 
dualised
\be\label{R6 in d1-7}
   \hat P_{1} = \epsilon_{1234567}R^{234567},
\ee
and obviously contains the time-like coordinate. Using the temporal 
involution on the $A_7\sim SL(8)$ algebra and using the momentum
generator to enhance the $SO(1,6)$ Lorentz algebra we find the full
local subgroup to be $SO(2,6)$. This time the generator 
\be
   K_c = \hat P_c + \eta_{cd} \hat P^d\qquad \eta_{ab}=(-,-,+,+,+,+,+,+)
\ee 
parametrises the coset $\frac{SO(2,6)}{SO(1,6)}$ which is well-known 
to be $AdS_7$. Using terminology of the conformal group we called 
this 'momentum generator' $K_a$ - the generator of special conformal 
transformations.\\

As a complete solution for the vacuum we find $S^4\times AdS_7$. 
This is a known result; this time, however, derived from a purely 
group theoretical point of view.

\subsection{Splitting $SL(11)$ into different $SL(d)$ subalgebras}

First of all we note that one can also consistently decompose 
$SL(11)\to SL(4)\times SL(7)$ with the latter tensor product exchanged 
relative to the previous discussion. This corresponds to the deletion of 
node 4 in {\it Figure A}. The roots of $E_{11}$ generating the {\bf 165}
representation ($R^3$) and the {\bf 462} representation ($R^6$) of 
$SL(11)$ can again be used to extend the obvious $SL(4)\times SL(7)$ 
subalgebra to $SL(5)\times SL(7)$ and $SL(4)\times SL(8)$ (but not 
simultaneously!). They are given respectively by: 
\begin{small}
\begin{eqnarray}
  SL(5)\times SL(7)&:& \lambda = \alpha_2 +2 \alpha_3+ 3\alpha_4 +
   3\alpha_5 +3\alpha_6+ 3\alpha_7 +3\alpha_8 +2\alpha_9
   +\alpha_{10} +\alpha_{11}\\
  SL(4)\times SL(8)&:& \lambda = \alpha_6+ 2\alpha_7 +3\alpha_8 
   +2\alpha_9 +\alpha_{10} +2\alpha_{11} 
\end{eqnarray}
\end{small}
Again, we notice that the field generated by the level 3 generator 
$R^{a_1\ldots a_8,b}$ can not survive on either subspace. Therefore 
the split Dynkin diagram shown in {\it Figure E} gives an effective
algebra, and we have used the generators $R^{a_1a_2a_3}$, and 
$R^{a_1\ldots a_6}$ to extend the relevant subalgebras exactly as 
discussed in the previous section.
\\[5ex]
\parbox{\textwidth}{
\begin{picture}(-150,0)(-50,30)

 \put(10,25){\line(0,1){10}}
 \put(10,40){\circle{10}}\put(6,47){$R^3$}

 \put(10,20){\circle{10}}\put(8,5){1}
 \put(15,20){\line(1,0){10}}
 \put(30,20){\circle{10}}\put(28,5){2}
 \put(35,20){\line(1,0){10}}
 \put(50,20){\circle{10}}\put(48,5){3}
 \put(90,20){\circle{10}}\put(88,5){5}
 \put(95,20){\line(1,0){10}}
 \put(110,20){\circle{10}}\put(108,5){6}
 \put(115,20){\line(1,0){10}}
 \put(130,20){\circle{10}}\put(128,5){7}
 \put(135,20){\line(1,0){10}}
 \put(150,20){\circle{10}}\put(148,5){8}
 \put(155,20){\line(1,0){10}}
 \put(170,20){\circle{10}}\put(168,5){9}
 \put(175,20){\line(1,0){10}}
 \put(190,20){\circle{10}}\put(188,5){10}

 \put(90,25){\line(0,1){10}}
 \put(90,40){\circle{10}}\put(86,47){$R^6$}
\end{picture}
}\\
\vspace{.5cm}\\
\centerline{\it Figure E}
\noindent
Again we can make the identification of the momentum generator with 
the generator of the 6-form potential. This time the relevant 
dualisation 
\be
   P_5 = \epsilon_{5\,6\,7\,8\,9\,10\,11} R^{6\,7\,8\,9\,10\,11} 
\ee
does not involve a time index (as opposing \eqref{R6 in d1-7}). The 
temporal involution just generates the subgroup $SO(8)$ which we would 
like to interpret as $SO(7)$ and the momentum generator. This momentum 
generator then parametrises the seven sphere $S^7$. The group covariant 
expression for 6-form generators (also group covariant with respect to 
the relevant conformal group) depends only on seven indices 
$5,\ldots,11$ and therefore has to be proportional to the totally 
antisymmetric tensor in this dimension. It is natural to identify the 
field strength with the volume form of the seven sphere $S^7$. 
\par\noindent
On the left hand side of the Dynkin diagram we find the 3-form potential 
to couple to the first node which involves the time direction. The 
interpretation of the simple positive root as a momentum generator via
\be
   \hat P_1 =\epsilon_{2341}R^{234}
\ee
implies that the invariant generators of the temporal involution span
the group $SO(2,3)$. However, as usually we interpret the additional
node as a momentum generator parametrising the coset $SO(2,3)/SO(1,3)$ 
which corresponds to $AdS_4$. This solution thereby corresponds to the 
other known vacuum solution of eleven dimensional supergravity 
$AdS_4\times S^7$.\\

Other cases to consider concern the splitting of $A_{10}\to A_5\oplus A_4$,
depicted in {\it Figure F}, where we have already enhanced the relevant
subalgebras by the representations antisymmetric in three and six indices
respectively, again making use of the fact, that the level 3 generator 
$R^{a_1\ldots a_8,b}$ can not couple to either side of the split.  
\\[5ex]
\parbox{\textwidth}{
\begin{picture}(-150,0)(-50,30)

 \put(10,40){\circle{10}}\put(6,47){$R^6$}

 \put(10,20){\circle{10}}\put(8,5){1}
 \put(15,20){\line(1,0){10}}
 \put(30,20){\circle{10}}\put(28,5){2}
 \put(35,20){\line(1,0){10}}
 \put(50,20){\circle{10}}\put(48,5){3}
 \put(55,20){\line(1,0){10}}
 \put(70,20){\circle{10}}\put(68,5){4}
 \put(75,20){\line(1,0){10}}
 \put(90,20){\circle{10}}\put(88,5){5}
 \put(130,20){\circle{10}}\put(128,5){7}
 \put(135,20){\line(1,0){10}}
 \put(150,20){\circle{10}}\put(148,5){8}
 \put(155,20){\line(1,0){10}}
 \put(170,20){\circle{10}}\put(168,5){9}
 \put(175,20){\line(1,0){10}}
 \put(190,20){\circle{10}}\put(188,5){10}

 \put(150,25){\line(0,1){10}}
 \put(150,40){\circle{10}}\put(146,47){$R^3$}
\end{picture}
}\\
\vspace{.5cm}\\
\centerline{\it Figure F}
\noindent
We find $R^{a_1\ldots a_6}$ not to be coupled to the 6 dimensional 
subspace to the left either. The extra scalar node (called $R_6$) can 
be seen to arrive by using the $\epsilon$ symbol of 6d space-time 
($A_5$) to contract all indices of $R^{a_1\ldots a_6}$. 
\par\noindent
This time we do not have available the interpretation of momentum 
generators which can only be used when the additional node couples
to the far left side of an $A_{d-1}$ subalgebra. In fact, we do not
find it on either side of this split. Another simple thought shows
that the Lorentz covariant and gauge invariant object of the theory 
-the field strengths- for example $F_{a_1\ldots a_7}$ has to vanish 
since the indices only range over 6 (or 5) different values. The eleven 
dimensional generalised self-duality condition (which is the only 
Lorentz covariant way of relating the group covariant objects belonging 
to dual generators) then automatically puts to zero the dual 
field strength $\partial_\mu \hat A_{a_1a_2a_3}$. We find that although 
the 6-form potential can live in a six dimensional subspace, its field 
strengths can not.
\par\noindent
In a very similar way also the discussion for the splitting 
$A_{10}\to A_2\oplus A_7$ goes wrong since one can not interpret
the gauge field nodes as 'momentum' generator. Also, the relevant
field strengths with antisymmetric indices are not compatible with
the dimensionality of the subspaces. The situation becomes 
even worse if one tries to place the 6-form potential in this $A_2$ 
subspace. Finally, it is easy to see that also $A_{10}\to A_1\oplus A_8$ 
can be excluded.

\section{Other supergravity theories}
If the above analysis reproduces the known results in case of eleven 
dimensional supergravity it is natural to ask whether it also does 
for the ten dimensional theories. We start considering the case of 
IIB theory. This theory was shown to also be in relation with a 
non-linear realisation of $E_{11}$ [2]. It might therefore be surprising 
that we will indeed find the known $S^5\times AdS_5$ solution for 
this case as well.
\par\noindent
The novel feature here is that the gravity subalgebra is $A_9$ instead
of the full $A_{10}$. It is natural to decompose the representations
of $E_{11}$ into those of $A_9$ in the way that is indicated in
{\it Figure G}, where we have placed the two expansion nodes
above the relevant gravity line.
\\[5ex]
\parbox{\textwidth}{
\begin{picture}(-150,0)(-50,30)


 \put(10,20){\circle{10}}\put(8,5){1}
 \put(15,20){\line(1,0){10}}
 \put(30,20){\circle{10}}\put(28,5){2}
 \put(35,20){\line(1,0){10}}
 \put(50,20){\circle{10}}\put(48,5){3}
 \put(55,20){\line(1,0){10}}
 \put(70,20){\circle{10}}\put(68,5){4}
 \put(75,20){\line(1,0){10}}
 \put(90,20){\circle{10}}\put(88,5){5}
 \put(95,20){\line(1,0){10}}
 \put(110,20){\circle{10}}\put(108,5){6}
 \put(115,20){\line(1,0){10}}
 \put(130,20){\circle{10}}\put(128,5){7}
 \put(135,20){\line(1,0){10}}
 \put(150,20){\circle{10}}\put(148,5){8}
 \put(155,20){\line(1,0){10}}
 \put(170,20){\circle{10}}\put(168,5){9}

 \put(150,25){\line(0,1){10}}
 \put(150,40){\circle{10}}\put(160,40){$B_2$}
 \put(150,45){\line(0,1){10}}
 \put(150,60){\circle{10}}\put(160,60){$\chi$}

\end{picture}
}\\
\vspace{.5cm}\\
\centerline{\it Figure G}
\noindent
The decomposition of representation has in great detail been carried out
in [5,6], where it has also been shown that very-extended $E_8$, also 
called $E_8^{+++}$ contains the subalgebra $E_7^{+++}$. Both algebras 
contain the same $A_9$ subalgebra but differ in the remaining nodes. We 
depict $E_7^{+++}$ in the following {\it Figure H}.
\\[5ex]
\parbox{\textwidth}{
\begin{picture}(-150,0)(-50,30)


 \put(10,20){\circle{10}}\put(8,5){1}
 \put(15,20){\line(1,0){10}}
 \put(30,20){\circle{10}}\put(28,5){2}
 \put(35,20){\line(1,0){10}}
 \put(50,20){\circle{10}}\put(48,5){3}
 \put(55,20){\line(1,0){10}}
 \put(70,20){\circle{10}}\put(68,5){4}
 \put(75,20){\line(1,0){10}}
 \put(90,20){\circle{10}}\put(88,5){5}
 \put(95,20){\line(1,0){10}}
 \put(110,20){\circle{10}}\put(108,5){6}
 \put(115,20){\line(1,0){10}}
 \put(130,20){\circle{10}}\put(128,5){7}
 \put(135,20){\line(1,0){10}}
 \put(150,20){\circle{10}}\put(148,5){8}
 \put(155,20){\line(1,0){10}}
 \put(170,20){\circle{10}}\put(168,5){9}

 \put(110,25){\line(0,1){10}}
 \put(110,40){\circle{10}}\put(106,47){$C_4$}
\end{picture}
}\\
\vspace{.5cm}\\
\centerline{\it Figure H}
\noindent
As indicated in {\it Figure H} the node apart from the gravity line
corresponds to a 4-form potential. In fact, splitting representation of 
$E_7^{+++}$ into representation of $A_9$ yields the fields
\be\label{E7+++ level 1 and 2}
   R^{a_1\ldots a_4},\quad {\rm and}\quad R^{a_1\ldots a_7,b}
\ee
at level 1 and 2 respectively. It can be shown that the relevant 
non-linear realisation describes the theory which is a consistent 
truncation of IIB supergravity, namely the one where all gauge
fields have been set to zero except the self-dual 5-form field 
strength [2,6]. 
However, we are now in the position to resume the discussion from 
chapter 2. We split the $A_9$ subalgebra (for good reason) into 
$A_4\oplus A_4$ by deleting the middle node 
$SL(10)\to SL(5)\times SL(5)$. The 4-form generator that normally 
takes values in all 10 dimensions can now be split into two 4-forms 
one of which takes values in dimensions $1,\ldots,5$ while the other 
takes values in dimensions $6,\ldots, 10$. We call these generators 
$R_4^+$ and $R_4^-$, where $R^+$ is the simple root of $E_7^{+++}$
while $R_4^-$ corresponds to the root
\be
   \alpha_{C_4} + \alpha_9 + 2\alpha_8+3\alpha_7 +4\alpha_6 +4\alpha_5
                + 3\alpha_4+ 2\alpha_3 + \alpha_2.
\ee
These roots can now be used to enhance the relevant $A_4$ subalgebras 
in the way indicated by {\it Figure H}. The algebra depicted in this 
figure is not a subalgebra of $E_7^{+++}$. However, it is an effective 
algebra describing the 10 dimensional theory since the field generated 
by the level 2 generator in \eqref{E7+++ level 1 and 2} does not respect 
the 5,5 split of ten dimensions and can in this particular case not be 
physical. As usually, we interpret that to provide a natural cut-off 
for {\it all} higher level generators as they would have to contain 
this non-physical field in a non-linear realisation.
\\[5ex]
\parbox{\textwidth}{
\begin{picture}(-150,0)(-50,30)


 \put(10,20){\circle{10}}\put(8,5){1}
 \put(15,20){\line(1,0){10}}
 \put(30,20){\circle{10}}\put(28,5){2}
 \put(35,20){\line(1,0){10}}
 \put(50,20){\circle{10}}\put(48,5){3}
 \put(55,20){\line(1,0){10}}
 \put(70,20){\circle{10}}\put(68,5){4}
 \put(110,20){\circle{10}}\put(108,5){6}
 \put(115,20){\line(1,0){10}}
 \put(130,20){\circle{10}}\put(128,5){7}
 \put(135,20){\line(1,0){10}}
 \put(150,20){\circle{10}}\put(148,5){8}
 \put(155,20){\line(1,0){10}}
 \put(170,20){\circle{10}}\put(168,5){9}

 \put(110,25){\line(0,1){10}}
 \put(110,40){\circle{10}}\put(106,47){$R_4^+$}
 \put(10,25){\line(0,1){10}}
 \put(10,40){\circle{10}}\put(6,47){$R_4^-$}

\end{picture}
}\\
\vspace{.5cm}\\
\centerline{\it Figure I}
\noindent
By this way of drawing, the potentials obviously depend only on the 
space-time coordinates they are attached to. As in chapter 2 we find it 
possible to interpret the nodes corresponding to $R_4^\pm$ as 
'momentum' generators. In applying exactly the same analysis as in
the previous chapter we find these momentum generators to parametrise
the coset spaces 
\be
  \frac{SO(2,4)}{SO(1,4)}\times \frac{SO(6)}{SO(5)}\sim AdS_5\times S^5.
\ee
We merely note that splitting the $A_9$ subalgebra in any other way
will not lead to consistent results.
\par
Performing a similar analysis on other very-extended groups 
$\mathcal{G}^{+++}$ gives analogous results. Let us briefly carry out 
the analysis for groups of the $D_n$-series of Lie algebras, i.e. we 
take the Kac-Moody groups $D_n^{+++}$. The discussion goes through for 
each value of $n$, but we will draw the diagram of $D_8^{+++}$ which was
also considered in [6].
\\[5ex]
\parbox{\textwidth}{
\begin{picture}(-150,0)(-50,30)


 \put(10,20){\circle{10}}\put(8,5){1}
 \put(15,20){\line(1,0){10}}
 \put(30,20){\circle{10}}\put(28,5){2}
 \put(35,20){\line(1,0){10}}
 \put(50,20){\circle{10}}\put(48,5){3}
 \put(55,20){\line(1,0){10}}
 \put(70,20){\circle{10}}\put(68,5){4}
 \put(75,20){\line(1,0){10}}
 \put(90,20){\circle{10}}\put(88,5){5}
 \put(95,20){\line(1,0){10}}
 \put(110,20){\circle{10}}\put(108,5){6}
 \put(115,20){\line(1,0){10}}
 \put(130,20){\circle{10}}\put(128,5){7}
 \put(135,20){\line(1,0){10}}
 \put(150,20){\circle{10}}\put(148,5){8}
 \put(155,20){\line(1,0){10}}
 \put(170,20){\circle{10}}\put(168,5){9}

 \put(150,25){\line(0,1){10}}
 \put(150,40){\circle{10}}\put(146,47){$R^2$}
 \put(70,25){\line(0,1){10}}
 \put(70,40){\circle{10}}\put(66,47){$R^6$}

\end{picture}
}\\
\vspace{.5cm}\\
\centerline{\it Figure J}
\noindent
At level 1, 2, and 3 one finds representations of antisymmetric 
tensors in 2, 6 and 8 indices, as well as a generators with 8 indices 
of which only 7 are antisymmetric and which we call $R^{7,1}$. This 
representation content shows that we find enhanced subalgebras of $A_9$ 
if we split $SL(10)\to SL(7)\times SL(3)$ (by deleting node $7$) or 
$SL(10)\to SL(3)\times SL(7)$ (by deleting node $3$). Repeating the same 
arguments as in chapter 2, and using the roots
\begin{eqnarray}
   \lambda_1 &=& \alpha_{R^2},\\
   \lambda_2 &=& \alpha_{R^6}+\alpha_9+ 2\alpha_8+3\alpha_7+3\alpha_6+
             3\alpha_5 + 3\alpha_4 + 2\alpha_3+ \alpha_2,
\end{eqnarray}
we find evidence for the existence of vacuum solutions $AdS_7\times S^3$.
Using the roots
\begin{eqnarray}
   \lambda_1 &=& \alpha_{R^6},\\
   \lambda_2 &=& \alpha_{R^2}+\alpha_9+ 2\alpha_8+2\alpha_7+2\alpha_6+
             2\alpha_5 + 2\alpha_4 + 2\alpha_3+ \alpha_2,
\end{eqnarray}
we also find the $AdS_3\times S^7$ solution. All other splits of $A_9$ 
will not result in appropriate cosets.
\par\noindent
In a similar fashion one shows that there does not appear to exist such
a vacuum solution for $E_6^{+++}$. This concludes the discussion for 
simply laced Kac-Moody algebras. It is easy to check that the only other 
possible $AdS\times S$ spectrum seems to occur for $G_2^{+++}$ with 
vacuum solutions $AdS_3\times S^2$ and $AdS_2\times S^3$.

\section{Conclusions}
We consider isometries of vacuum solutions of supergravity theories
and identify their algebras within a special class of Kac-Moody 
algebras, namely the so-called very-extended Lie algebras. The 
low level generators of these algebras were shown to describe 
(super)gravity theories in a non-linear realisation. We discover the 
appropriate isometry algebras after solving a consistency puzzle.
One cannot derive the isometry algebras of these vacuum solutions 
canonically when starting from any other than the very-extended algebras. 
It is an essential feature of our discussion that the generators 
corresponding to duals of gravity naturally vanish. By using the 
temporal involution [8] we furthermore found the right signatures 
of the subgroups to turn up in the correct way. It would be interesting 
to discover less simple solutions of 11d supergravity via a similar 
approach. \\
Amazingly all the solutions found so far support supersymmetry. In [12] 
the relation of coset spaces to supersymmetry was explained and it was 
shown that the Killing spinor equation can be understood geometrically.
It would be challenging to see how supersymmetry can be embedded into 
the full Kac-Moody algebra $E_{11}$.

\section{Acknowledgments}
IS is grateful for talks with D. Kazhdan, F. Englert, and P. West. 
The work of AM is supported by the European Commission RTN programme 
HPRN-CT-2000-00131 and the Deutsche Forschungs\-ge\-meinschaft (DFG). The work of IS is partly supported by 
BSF - American-Israel Bi-National Science Foundation, the Israel 
Academy of Sciences - Centers of Excellence Program, the 
German-Israel Bi-National Science Foundation, and the European 
RTN-network HPRN-CT-2000-00122.

\end{document}